# Deep sea floor observations of typhoon driven enhanced ocean turbulence

by Hans van Haren[1*], Wu-Cheng Chi[2], Chu-Fang Yang[2], Yiing-Jang Yang[3], Sen Jan[3]

[1]Royal Netherlands Institute for Sea Research (NIOZ) and Utrecht University, P.O. Box 59, 1790 AB Den Burg, the Netherlands.
*e-mail: hans.van.haren@nioz.nl
[2]Institute of Earth Sciences, Academia Sinica, No.128 Academia Road, Sec 2, Nangang Dist., Taipei, Taiwan, R.O.C
[3]Institute of Oceanography, National Taiwan University, No. 1, Sec. 4, Roosevelt Road, Taipei 10617, Taiwan, R.O.C.

ABSTRACT

The impact of large atmospheric disturbances on deep benthic communities is not well known quantitatively. Observations are scarce but may reveal specific processes leading to turbulent disturbances. Here, we present high-resolution deep-ocean observations to study potential turbulent mixing by a large atmospheric disturbance. We deployed an array of 100-Hz sampling-rate geophysical broadband Ocean Bottom Seismometers (OBSs) on the seafloor. Within the footprint of this array we also deployed an oceanographic 0.5-Hz sampling-rate vertical temperature sensor string covering the water phase between 7 and 207 m above the seafloor at about 3000 m depth off eastern Taiwan between June 2017 and April 2018. In September 2017, all instruments recorded Category 4 cyclone Typhoon Talim's passage northeast of the array one day ahead of the cyclone's closest approach when the cyclone's eye was more than 1,000 km away. For 10 days, a group of near-inertial motions appeared most clearly in temperature. The group contained the largest inertial amplitudes in the ten month time series, and which led to turbulence dissipation rates $O(10^{-7}$ $m^2$ $s^{-3})$. The observation reflects the importance of barotropic response to a cyclone and the propagation of inertio-gravity waves in weak density stratification. In addition to internal tides, these waves drove turbulent overturns larger than 200 m that were concurrently recorded by OBSs. The turbulent signals were neither due to seismic activity nor to ocean-surface wave action. Cyclones can generate not only microseisms and earth hums, as well as turbulence in the water column, producing additional ground motions. Quantified turbulence processes may help constrain models on sediment resuspension and its effect on deep-sea benthic life.

## 1. Introduction

It is well known that large energetic atmospheric disturbances such as tropical cyclones not only affect the Earth's surface both onshore and offshore, but also the ocean interior (e.g., D'Asaro et al., 2014; Jan et al., 2017). One means of transporting the energy of cyclones downward into the ocean interior is by surface wind-wave action, which reaches deeper depths when wavelengths are longer. For large cyclones, such propagating waves over deep water may be felt hundreds of meters from the surface, whereby they can reach near-coastal continental shelf bottoms and affect living organisms there (e.g., Sanchez-Vidal et al., 2012). However, as the energy and orbital motions of propagating surface wind waves decrease exponentially with depth, it is not known what their impact is on the deep-ocean floor at more than 1000 m depth.

On the Earth, which rotates with angular velocity **Ω**, the passage of atmospheric disturbances will also generate larger periodic oscillatory near-inertial transient motions in the ocean after geostrophic adjustment (Gill, 1982). These motions have inertial periods that depend on latitude $\varphi$.

The inertial frequency $f = 2\Omega \sin\varphi$, $\Omega = |\mathbf{\Omega}|$, of about 1 cpd, cycles per day, at $|\varphi| \approx 30°$, is the lowest frequency of freely propagating internal waves that are supported by the stable density $\rho(T,S)$ stratification $-d\rho/dz$ (z is the vertical coordinate, T denotes temperature, and S denotes salinity). The ocean is stably stratified mainly due to solar heating from above, which contrasts with daytime atmospheric heating from below. Buoyancy frequency $N = \{-g/\rho(d\rho/dz + g\rho/c_s^2)\}^{1/2}$ (e.g., Gill, 1982) $\approx 10^{-3}$-$10^{-2}$s$^{-1}$ $\approx$ 10-100 cpd (g denoting the acceleration of gravity and $c_s$ the speed of sound) sets the upper frequency bound for internal waves. As a result, the frequency band $\sigma \in [f, N]$ contains internal wave motions near f that are near-horizontal and motions near N that are near-vertical. Such motions include internal tides whose energy propagates along slopes with angles $\beta(\sigma, f, N) = \cos^{-1}[((\sigma^2-f^2)/(N^2-f^2))^{1/2}]$ to the horizontal direction (e.g., LeBlond and Mysak, 1978). Hence, in a well-stratified open ocean

near-inertial motions are more notable in the horizontal current meters than T,S-measurements. The opposite is true for near-buoyancy frequency motions.

A rare exception occurs when stratification is negligible, $N < 2\Omega$. In that case, gravity is no longer the dominant restoring force driving the internal waves supported by the density stratification, but variations in the Earth's angular momentum support 'gyroscopic' waves (LeBlond and Mysak, 1978). Gyroscopic waves are generated as a disturbance in the balance between the centrifugal force and the pressure gradient force, so that in an inertial frame of reference the Coriolis force acts as restoring force. Only internal waves at f can propagate freely from purely homogeneous to vertically stratified waters locally in all horizontal directions. Gyroscopic motions follow a strictly circular path in their plane of propagation, which orients relative to $\Omega$ as a function of frequency. Except at the North Pole where an inertial gyroscopic motion would be perfectly circular, a horizontal current meter records this circular gyroscopic motion on a plane perpendicular to gravity as an ellipse, of which the ratio of minor-to-major axis is a function of latitude. As a result, such motion at sub-tropical latitudes may be found as strongly elliptic horizontal current velocities ($u^f$, $v^f$) (van Haren and Millot, 2004), the super-script indicating the frequency pass-band, as well as non-negligible $T^f$ and vertical current $w^f$, and possibly pressure $p^f$.

Near the ocean surface in relatively strong stratification, mainly near-circular horizontal motions are generated at f, of which the energy of baroclinic waves slowly radiates downwards (cf., near-surface measurements in Cuypers et al., 2013; D'Asaro et al., 2014). After about a week or longer, surface-generated baroclinic inertial wave energy could reach the deep-ocean floor, whereby their progressively steeper vertical component is due to decreasing f as they move equatorward and, in addition, by N decreasing with depth when moving downward (Garrett, 2001). Near-inertial waves can freely propagate only a small distance poleward from their initial latitude before reflecting equatorward. Atmospheric tropical storms generate a band of inertial motions because of the finite latitudinal extent of a storm. Alternatively, cyclonic vorticity $\zeta > 0$ in an environment of mesoscale sub-inertial motions may cause local kinetic energy to peak at slightly super-inertial frequencies, so that the internal wave motions experience an effective total vorticity $f_{eff} = f + \zeta/2 > f$ upon generation (Kunze, 1985).

However, current meter observations have also shown immediate near-inertial responses at depths exceeding 1000 m within a few hours around the passage of a tropical cyclone (Shay and Elsberry, 1987; Morozov and Velarde, 2008; Zhang, 2013). Such a fast response was attributed to barotropic surface displacements, which are part of the geostrophic adjustment solution (Rossby, 1938; Gill, 1982; Chang, 1985). A fast near-inertial response is also typical in shallow seas, which occurs via a lateral boundary condition so that inertial motions appear as (quasi-)mode-1 between surface and bottom boundary layers, with a slower baroclinic response locally (Millot and Crépon, 1981; Maas and van Haren, 1987).

Near-inertial motions have relatively short vertical scales, so that they dominate the excitation of vertical current shear in the stratified ocean (e.g., LeBlond and Mysak, 1978). Thus, they are vital for creating turbulent diapycnal overturning in the form of Kelvin-Helmholtz instabilities, possibly by disturbing high-frequency internal waves. A group of cyclone-generated near-inertial waves may greatly contribute to ocean-interior turbulence.

In this study, we are interested in a tropical cyclone's potential impact on the deep ocean near slope bathymetry. In particular, we study the relation between internal waves in the water phase and ground motions on the seafloor. Sediment resuspension may occur via bed shear stress of horizontal currents, but also, probably more importantly, via sheared internal wave breaking (Eriksen, 1982; Thorpe, 1987) and turbulent convection. The hypothesis is that near-inertial waves generate nonlinear internal waves that may break in the interior but foremost when they interact with the deep sloping seafloor. We test this hypothesis by using a combination of a broadband Ocean Bottom Seismometer (OBS) array, commonly used for earthquake observations but capable of observing ocean turbulence and ocean surface waves (e.g., Webb, 1998; Chi et al., 2010a; Chi et al., 2010b; Wang et al., 2010; van Haren, 2011; Davy et al. 2014), and oceanographic densely spaced high-resolution temperature sensors, which monitors internal-wave-induced turbulence (van Haren and Gostiaux, 2012). We do not focus on the frequency band centered around 0.2 Hz that includes "microseisms" from surface ocean standing waves but on the frequency band centered around 0.02 Hz that includes very long "infra-gravity" propagating surface wind waves (Webb, 1998) and ocean interior turbulence (D'Asaro and Lien, 2000; van Haren, 2011) and on the inertial band of inertio-gravity waves (IGW; internal waves propagating under the influence of buoyancy and Coriolis forces). The study area, east of Green Island, Taiwan, is one of the world's most typhoon-affected regions (cf., NASA 150 years of tropical cyclone tracks, intensity and frequencies map until September 2006 available from https://earthobservatory.nasa.gov/images/7079/historic-tropical-cyclone-tracks and NOAA's National Hurricane Center website https://www.nhc.noaa.gov/, with tracking format by Landsea and Franklin, 2013).

## 2. Materials and methods

*2.1. General site characteristics*

Between the summer of 2017 and the spring of 2018, we deployed mooring instruments east of Taiwan on the relatively steep underwater slopes of the volcano Green island (Figs 1, 2). We collected data from three OBSs (named A, B, and C in Fig. 2) and a temperature T-sensor string (named T). The T-sensors mooring was deployed on 9 June 2017, which is yearday 159, and the OBSs were deployed on 29-30 August, which is yearday 240-241. Here we use the convention that January 1, 12:00 UTC = yearday 0.5. All instrumentation was recovered on 20 April 2018, which is yearday 109+365 = 474. Locally, the bottom slope $\gamma$ = 10±5°, on 100 m scales (Fig. 1). This bottom slope is roughly the same as the slope for propagating semidiurnal internal tides $\beta(M_2) \approx \gamma$. The bottom slope is supercritical for propagating near-inertial waves. The mooring sites were closely spaced, typically 500 m apart (Fig. 2), in accordance with the expected short horizontal decorrelation scales of underwater internal waves and turbulence. At half coherence, the scales are a few km at 2 cpd of internal tides (Briscoe, 1975) and about 10 m at 100 cpd of small-scale internal waves and larger energetic turbulent overturning (van Haren et al., 2016). The central coordinates of the mooring array were at 22° 39.7´N, 121° 36.2´E, H = 3150 m water depth, which varied by several 100 m between the individual sites. The local inertial frequency (f) is $5.6\times10^{-5}$ $s^{-1}$ which is approximately 0.77 cpd.

*2.2. Array of Ocean Bottom Seismometers*

The OBSs were equipped with Trillium Compact 120 broadband velocity sensors that have a flat response between 0.0083 and 100 Hz. The lower and higher frequency signals outside the flat response can be amplified after being corrected for the instrument response. Each OBS also had a differential pressure gauge (Webb, 1998). The OBS uses an in-house digitizer with a sampling rate of 100 Hz. The sensor is housed in a sphere with a gimbal system that will level the sensor periodically in a prescribed time interval.

The purpose of the OBS is to record earth motions in three Cartesian axes X, Y (both horizontal) and Z (vertical), in addition to pressure (p) perturbation. OBS-X and OBS-Y are the ground motions of the east-west and north-south components, respectively. This geographic orientation was corrected by using the reference from polarization of airgun-shot waveforms. Pressure can also record surface water waves at great depths, including microseisms and infra-gravity waves (Webb, 1998). In the infra-gravity wave band, turbulence induced by internal wave breaking is also expected (D'Asaro and Lien, 2000; van Haren, 2011). For seismic waves propagating in the solid earth due to earthquakes, the OBSs will record ground motions almost simultaneously because of the short distance of a few 100 m between the instruments (Fig. 2) and the fast medium velocity in the crust. Storm-generated long surface ocean waves have propagation speeds of about 200 m $s^{-1}$ in the open ocean. The array should variably record these surface waves almost simultaneously depending on their wavelengths, due to the lower frequency bands we used for this study. Turbulence due to internal wave breaking is expected to vary over horizontal distances of about 100 m, and may thus vary substantially between the OBSs. We are unaware of previous investigations of turbulence in OBS-X and OBS-Y records, in which we are especially interested.

From the raw OBS data, pressure component OBS-p was derived from the factory-provided instrument response correction information and by using a band-pass filter with cut-off frequencies at 0.4 and $10^3$ cpd. The filter-pass band potentially includes semidiurnal tidal and inertial motions. The signals below 288 cpd (0.0033 Hz) in OBS-Z and OBS-p are sensitive and related to environmental temperature perturbation because of insulation configurations of the instrument (Cox et al., 1984; Doody et al., 2018). As a result, the low-frequency temperature-related signals recorded by the OBS array can be used for directional analysis of internal waves. The horizontal OBS-X and OBS-Y were recomputed to a velocity signal by band-pass filtering between $10^2$ and $10^3$ cpd, so that seismic and microseism signals were effectively reduced.

*2.3. Temperature sensors mooring*

The T-sensor mooring covered a range between 7 and 207 m above the seafloor (z = -2937 and -3137 m, respectively). A 0.0063 m diameter nylon-coated steel cable held 101 'NIOZ4' self-contained temperature sensors at 2-m intervals (van Haren et al., 2009). The sensors sampled at a rate of 0.5 Hz and their clocks were synchronized to within 0.02 s via induction every 4 h. The sensors' precision is less than 0.0005°C; the noise level less than 0.0001 °C. Electronic (noise, battery) problems failed 13 sensors in the first half-year and their data were linearly interpolated from neighboring sensors.

A low-resolution, single-point Nortek AquaDopp acoustic current meter was deployed at 208 m above the seafloor (z = -2936 m). It sampled at a rate of 0.0016 Hz. It was below 14 Benthos glass spheres that provided 250 kg net buoyancy. Since the mooring site was underneath the general path of the Kuroshio current, care was taken to minimize displacements by current drag. The net buoyancy and thin cables ensured that vertical top-buoy-motions were less than 0.1 m typically, and less than 0.4 m maximum, under mean (maximum) current speeds of 0.15 (0.25) m s$^{-1}$. These displacement values were well less than half the distance between T-sensors.

*2.4. Other data sets*

For this work, we also used previously collected Multi-Beam data, gridded to 100x100 m resolution for bathymetry and for 2D bottom-slope determination. In addition, one shipborne SeaBird 911 Conductivity Temperature Depth CTD-profile was obtained after mooring recovery to within 10 m from the bottom and 1 km from the mooring site. The CTD-data were used to establish a temperature-density relationship to quantify turbulence parameter estimates from the moored T-sensor data.

The moored T-sensor data were converted to Conservative Temperature ($\Theta$) values (IOC, SCOR, IAPSO, 2010) and drift-corrected against a linearized profile before being used as a proxy for potential density anomaly $\sigma_3$ referenced to a level of 3000 dBar following the tight temperature-density anomaly relationship $\delta\sigma_3 = \alpha\delta\Theta$, $\alpha = -0.2083\pm0.0005$ kg m$^{-3}$°C$^{-1}$, which includes salinity contributions under local conditions (Fig. 3). The linear temperature-density relationship was the mean for the lower 500 m above the seafloor of the CTD-data.

*2.5. Ocean turbulence computation*

Turbulence was quantified using the analysis method by Thorpe (1977) on moored T-sensor data with temperature as tracer for density variations. Turbulent kinetic energy dissipation rate $\varepsilon$, which is proportional to turbulent diapycnal flux, and vertical eddy diffusivity $K_z$ were determined by calculating overturning scales. These scales were obtained after reordering every time-step the potential density (temperature) profile $\sigma_3(z)$, which may contain inversions, into a stable monotonic profile $\sigma_3(z_s)$ without inversions (Thorpe, 1977). After comparing raw and reordered profiles, displacements d = min(|z-z$_s$|)·sgn(z-z$_s$) were calculated that generated the stable profile. Then,

$$\varepsilon = 0.64 d^2 N^3, \tag{1}$$

where N denotes the buoyancy frequency computed from the reordered profile and the constant follows from empirically relating the overturning scale with the Ozmidov scale $L_O =$ $(\varepsilon/N^3)^{1/2}$: $(L_O/d)_{rms}$ = 0.8 (Dillon, 1982), a mean coefficient value from many data points distributed over an order of magnitude. Using $K_z = \Gamma\varepsilon N^{-2}$ and a mean mixing efficiency coefficient of $\Gamma$ = 0.2 for the conversion of kinetic into potential energy for high bulk Reynolds number values Re = $10^5$-$10^7$ ocean observations (not necessarily for low Re < $10^4$ laboratory observations) (Osborn, 1980; Oakey, 1982; Gregg et al., 2018), we find,

$$K_z = 0.128d^2N. \qquad (2)$$

It is noted that the mixing efficiency coefficient is a mean value of a distribution of a statistically large number of realizations an order of magnitude wide. While recent suggestions are given to better parameterize mixing efficiency using Froude number in Direct Numerical Simulations (Garanaik and Venayagamoorthy, 2019) or Ozmidov/Thorpe overturning scale ratio from shipborne microstructure data (Ijichi and Hibiya, 2018), the large number statistics from moored high-resolution temperature data above a deep-ocean steep slope as employed here demonstrated a collapse of Ellison and Thorpe scales to comparable mean value (Cimatoribus et al., 2014).

According to Thorpe (1977), results from (1) and (2) are only useful after averaging over the size of an overturn. Due to higher temperature precision and less artificial motion of the T-sensor moorings, compared with shipborne CTD-profiling data, we have smaller thresholds for genuine overturn discrimination from artificial and noise motions than some previous studies, e.g., Galbraith and Kelley (1996). The T-sensor threshold limits mean turbulence values to $<\varepsilon>_{thres} = O(10^{-12})$ m$^2$ s$^{-3}$ and to $<K_z>_{thres} = O(10^{-6})$ m$^2$ s$^{-1}$ in weakly stratified waters. Also, every 2 s an entire 200 m vertical profile is measured, which allows for averaging over many more data points than in a slanted CTD-profile. Here, averaging is applied over at least vertical scales of the largest overturns and over at least buoyancy time scales to warrant a concise mixture of convective- and shear-induced turbulence, and to justify the use of the above mean coefficient values. For the averaging, we replaced values below threshold with zeros. Compared with multiple shipborne shear- and temperature-variance profiling, the present method yields similar results within a factor of two (van Haren and Gostiaux, 2012).

## 3. Observations

Taiwan is hit by on average 3.7 typhoons per year (Wu and Kuo, 1999). The year of 2017 was a moderate typhoon season in the western North Pacific. Taiwan, and Green Island in particular, were largely missed and partially hit by 2 typhoons only, resulting in a considerable less water supply as typhoons can dump 1000 mm of rain (Wu and Kuo, 1999; Hung and Shih, 2019). On 5-6 September 2017, which is yearday 247-248, a modest Storm Guchol passed through the Luzon Strait equatorward of the mooring with wind speeds up to 18 m s$^{-1}$. Subsequently, the closest encounter during the mooring period was of Category 4 typhoon Typhoon Talim on 14 September, which is yearday 256, poleward of the instrumented site (Fig. 1). Typhoon Talim's eye was still more than 600 km away from the mooring site on that day. Also its outer circulation was never above the moorings. In the present paper, we focus our data analysis on this period after some general comments on the entire record. Further work on other periods will be presented in future reports.

*3.1. Overview*

The observed typical ocean current amplitudes of 0.15 m s$^{-1}$ were modulated at sub-inertial timescales O(100 days), with modulations at timescales O(10 days), at inertial (1.29 days) and at semidiurnal M$_2$-tidal (0.52 days) scales (Fig. 4a,b). The longer periodicity was reflected in T-variations (Fig. 4b), with generally warmer waters observed during large-scale southward flow. The sub-inertial variations were thought to be linked with the overlying Kuroshio current and with local coastal boundary effects. The f- and M$_2$-variations were associated with IGW. They are unlikely related with topographic Rossby waves, as N·tan($\gamma$) ≈ 9.5±1×10$^{-5}$ s$^{-1}$, which is about twice the local inertial frequency and distinctly different from f and M$_2$ for long-term, larger than inertial period, forcing and averaging. This independence of wavenumber for the topographic Rossby wave dispersion relation holds for the short-wave limit considered here, whereas for waves of the size of the Rossby scale and larger wavenumber dependence is anticipated (LeBlond and Mysak, 1978; Olbers et al., 2012).

*3.2. A large inertial motion event*

Generally, near-inertial motions were rather small compared to the total current and temperature amplitudes (Fig. 4a-c). However, at day 254.5±0.05 (12 September), T-variations occurred suddenly with large amplitudes seen as near-inertial oscillations $T^f$ that contributed more than 50% to the total temperature variations and which lasted for 10 days (Fig. 4d). During days 254-264, T(f) were at least twice those of other periods, including during tropical Storm Guchol. These were the largest $T^f$-amplitudes in the ten month mooring time series (Fig. 4c). The enhanced $T^f$ were not due to advection of fronts by horizontal near-inertial currents because these were weak at the time, but they were attributed to locally generated large near-inertial IGW, possibly after initial barotropic generation (see Section 4). In $u^f$ (and $v^f$) were less distinct, albeit the largest in the record, but they peaked about four days later around day 259.

The sudden increase in $T^f$ occurred about one day before the closest passage of Typhoon Talim north of the mooring when the eye of the cyclone was more than 1000 km away. Increase in inertial amplitude was also seen in low-frequency OBS-p between day 254.5 and about day 261 (Fig. 4e). All three OBSs recorded the increase in p(f), with slight variations between them. Both $T^f$ and $p^f$ demonstrated a reinforcement of their amplitude between days 258 and 261, when a factor of two enhancement in daily-mean turbulence values was observed (Fig. 4f, note the logarithmic vertical-axis scale). During these three days, wide-elliptic horizontal currents were observed at the mooring site, while after day 261 the currents were more rectilinear and no longer showing elliptic patterns associated with near-inertial motions. The major current-ellipse axis veered from WNW-ESE to SW-NE during Typhoon Talim's passage, suggesting particle motions in the WNW-ESE direction while the near-inertial wave energy was propagating in the perpendicular direction of NNE-SSW, see Fig. 5a,b.

The northerly direction is consistent with results from frequency-wavenumber analysis of low-frequency OBS-Z data between days 254 and 262 (Fig. 5c). The near-inertial waves were coming from the north to north-east of the array, focusing on 25 back-azimuth of the array. The waves had a mean slowness between 2000 and 11000 s km$^{-1}$, which gives the upper bound of the mean propagation speeds between 0.1 and 0.5 m s$^{-1}$ that are within the order of magnitude for baroclinic waves. The derived propagation speeds might be over-estimated if the normal of wave front passing though the OBS array is not horizontal.

The dominance of near-inertial motions in especially T(t) during the period between days 254 and 264 is appreciated in 10-day mean spectra (compare Fig. 6a with Fig. 6b). The clear super-inertial peak near f in T-variance in Fig. 6b is accompanied by a smaller peak at a frequency just higher than 2f (and higher harmonics), not at $M_2$. The inertial-harmonic peaks reflect the sudden appearance of a nonlinear near-inertial wave package. $M_2$-motions peak in kinetic energy spectra.

*3.3. Magnification of a large inertial motion event*

Figures 7 and 8 demonstrate in three-day detail the different effects of turbulent motions in moored OBS- and T-sensor data. Horizontal amplitudes OBS-C = (OBS-X$^2$ + OBS-Y$^2$)$^{1/2}$ had different appearances at different sites. It is noted in Fig. 7a that a coherent spike at day 257.28 results from an earthquake that occurred nearly 3 days after the onset of enhanced T$^f$. The apparent long duration of such earthquake signals are the result of the applied band-pass filter.

During Typhoon Talim's northward veering, between days 257 and 261, OBS-C (Fig. 7a) showed a modulation of enhanced activity more or less during the relatively warm phases of inertial motions, with smaller activity at the transition from the warming to the cooling phase (see T-data in Fig. 7b). The 200-m-tall T-range was not high enough to cover the amplitude of the near-inertial waves, which appeared quite vertical. The range was also not capable of resolving the largest overturns, which were seen to reach the lowest sensor close to the seafloor. The large overturns strained enhanced stratification in thin layers leaving >100 m high layers of N < 2Ω for an hour (Fig. 7c). The thin-layer stratification followed the large near-inertial amplitudes vertically. Also we cannot resolve the largest dissipation values at the upper or lower boundaries of the vertical range of the overturns. 200-m Vertically averaged dissipation rates varied over more than three orders of magnitude, having an approximate periodicity of half an inertial period (Fig. 7d). The tallest overturns lasted in duration similar to the buoyancy period of 12,000±3000 s (about 3.5 hours). Sometimes the overturns had shorter duration. It was noted that the near-inertial modulation in OBS-C did not always coincide with the relatively large interior non-bottom reaching turbulence during the warming phase. A reasonable correspondence was found between OBS-C and turbulence dissipation rate after day 258 (compare Fig. 7a and 7d). Also, the different timing in OBS-C activity at 500-m horizontal footprint of this seismic array might well be attributable to typical baroclinic phase speeds $O(0.1)$ m s$^{-1}$, as well as to unknown development of turbulent overturns. To study such development, one needs moorings at about 10 m horizontal scales, which were not employed here. The 3-day, 200-m mean turbulence values are for Fig. 7 $[<\varepsilon>] = 1.1\pm0.5\times10^{-7}$ m$^2$ s$^{-3}$, $[<K_z>] = 5\pm2\times10^{-2}$ m$^2$ s$^{-1}$, while $<[N]> = 5\pm1\times10^{-4}$ s$^{-1}$.

*3.4. Magnification of other three-day periods*

Tall overturns were also observed during other periods (e.g., the week before Typhoon Talim's passage, Fig. 8), but less (or not) dominant at the inertial frequency, and more often at smaller vertical scales. Semidiurnal tidal and higher frequency periodicities and modulations seemed more dominant, while the mean turbulence parameter and OBS-C values were about a factor of two smaller than in Fig. 7. During the period of Fig. 8, internal tide breaking seemed to be the dominant process causing turbulent and ground motion activity in the infra-gravity wave band. The 3-day, 200-m mean turbulence values, $[<\varepsilon>] = 0.4\pm0.2\times10^{-7}$ m$^2$ s$^{-3}$ $[<K_z>] = 2\pm1\times10^{-2}$ m$^2$ s$^{-1}$ while $<[N]> = 6\pm1\times10^{-4}$ s$^{-1}$, are less than half those for Fig. 7. Root-mean square overturn sizes and, hence, the Ozmidov scale are about half those of Fig. 7. During other periods, yearday 246-248, a complex of diurnal and semidiurnal internal tidal motions may dominate (Appendix A).

*3.5. A comparison in large magnifications*

In further magnifications, the difference between near-inertial (Fig. 9a) and tidal and higher frequency (Fig. 9b) turbulent internal wave motions were seen. When near-inertial IGW dominated (Fig. 9a), turbulent overturns were more similar to vertical column-like convection dominated on the large scale. When tidal motions dominated (Fig. 9b), turbulent overturns were more shear-induced across the stronger stratification, which was 4 times larger in Fig. 9b compared to in Fig. 9a. This may be inferred as follows.

The larger stratification is observed from counting the number of isotherms over the vertical range in Fig. 9. Stronger stratification distribution is also seen in the buoyancy frequency of Fig. 8c, in comparison with more vertically oriented alignments in Fig. 7c. In general, larger stratification can support larger shear for the same level of gradient Richardson number stability. It is noted however, that differences between various periods are subtle and a combination of the two types of convective and shear-induced turbulent overturning was seen in both examples of Figs 7 and 8, although perhaps at different scales. Such a combination is common in the oceanic high-Reynolds number turbulence with shear-induced motions generating secondary convection and vice-versa (e.g., Matsumoto and Hoshino, 2004). In both three-day periods of Figs 7 and 8, the turbulence was vigorous with overturns of cooler (denser) waters above warmer (less dense) waters occurring at various scales.

*3.6. Mean internal wave-turbulence spectra*

Five-month mean temperature spectra show a dominant and unusually large and coherent near-inertial peak, with more variance than in the semidiurnal tidal peak (Fig. 10). In stronger-stratified open-ocean waters near-inertial motions are considered to be mainly horizontal and do not show as a prominent peak in T-spectra (Pinkel, 1981; van Haren and Gostiaux, 2009). There, the vertical scale of semidiurnal internal tides is considerably larger than that at f, as was inferred from vertical coherence tests. In the present variance spectrum (Fig. 10a), the rest of the IGW-band is rather featureless and roughly scales with a power law of $\sigma^{-2}$, generally considered as the canonical internal wave slope (on a log-log plot). Around

N, the slope-scaling changes and in steps it reaches a slope-scaling of $\sigma^{-5/3}$ around $N_{max}$, which implies a mean dominance of shear-induced turbulence or a passive scalar (Tennekes and Lumley, 1972; Warhaft, 2000). $N_{max}$ is the maximum thin(2-m)-layer stratification.

The transition between N and $N_{max}$ does not have well-defined scaling, which reflects a mix of convective (active scalar) and shear-induced turbulent and small-scale internal wave motions.

The $\sigma^{-5/3}$-slope-scaling of the 'inertial subrange' is well-established and the spectrum statistically smooth for $\sigma > N_{max}$ until roll-off into noise (Fig. 10a). The $N_{max}$ is also a transition in coherence between all possible pairs of T-sensors at the vertical distance $\Delta z = 10$ m (Fig. 10b). At larger vertical distances, the coherent frequency range shrinks to lower frequencies, with a cut-off at about N for $\Delta z = 50\text{-}100$ m. At these and larger vertical distances, coherence is observed to be consistently higher at f, and even at 2f, than at $M_2$. This indicates that in the present data the vertical scale of inertial motions is larger than that of tidal motions, which contrasts with open-ocean-interior observations where the coherence peak was found at $M_2$ (van Haren and Gostiaux, 2009). For comparison, the ocean-interior data had N = 26 cpd, the $\Delta z = 10$ m coherence transition into noise was at 40 cpd and $N_{max}$ was at 80 cpd. In the present data, the value of N is about three times smaller and that of $N_{max}$ about 1.3 times smaller, while the transition frequency into noise for $\Delta z = 10$ m is about three times larger.

Small coherence peaks are seen around 4f, 6f and N (Fig. 10b), especially at separation scales >50 m. At $\Delta z = 180$ m, the coherence is still considerable at f and 2f, but approaches the noise level at $M_2$, which is close to the two-hour and 100-m averaged minimum buoyancy frequency $N_{min} \approx 2$ cpd $\approx 2\Omega$. The coherent near-inertial motion associates with large, 200 m tall variations in isotherms, which are not common in the stratified ocean and basically can only exist in weakly stratified waters.

## 4. Concluding discussion

The calculated turbulence dissipation rates $O(10^{-7})$ $m^2s^{-3}$ are comparable, and are within one standard deviation or less than a factor of two, with those from recent observations of intensely breaking internal waves above generally less deep sloping topography where freely propagating semidiurnal internal tides dominated (e.g., van Haren and Gostiaux, 2012). This supports the previous notion that boundary mixing is important and sufficient to maintain the ocean stratification and transport of nutrients and suspended matter. In the deep-ocean, turbulent overturning exceeds 100 m vertical scales extending from the seafloor, is much larger than frictional flow turbulence and is doubled when inertial internal wave breaking adds to internal tide breaking.

The present observations were made above topography about halfway between the coast and the abyssal plain, with small 1-km-scale ridges of variable seafloor slopes. The mooring array was above a slope that was approximately critical for internal tide propagation. However, within 1 km a super-critical slope could be found for semidiurnal internal tides (Fig. 1). Boundary-current-driven variations cause time-varying N and thus time-varying IGW-slopes. The seafloor-slope was always supercritical for propagating near-inertial waves. From the perspective of internal wave breaking at a supercritical slope, the mooring site is thus expected to be partially affected by semidiurnal internal tides, when stratification is favorable, and partially by near-inertial waves, when they occur.

Turbulent motions were modified and dissipation rates more than doubled above the internal tide-generated values during a period of 10 days around Typhoon Talim's passage to the north of the mooring. In particular, the cyclone, with a radius larger than 500 km at the time, was sensed at the deep boundary when its eye was 1000 km away. Typhoon Talim's approach is suggested from the initiation of inertial motions, with associated turbulence, and infra-gravity waves. Adopting the inertial wave motions as long wavelength shallow water surface waves, they travel about three to four times faster than Typhoon Talim's translation speed between days 254 and 256. This provides additional information to microseism observations associated with the passage of a tropical cyclone (Chi et al., 2010a; Chi et al., 2010b; Davy et al., 2014).

Our observations are consistent with previously reported observations of a rapid barotropic surface gradient-driven response, via a lateral boundary, of deep near-inertial motions (e.g., Morozov and Velarde, 2008). Four days after the first barotropic inertial response, the baroclinic near-inertial wave response occurred in our data, also noticed in current observations besides in temperature and OBS. Near-inertial waves can only propagate equatorward from their latitudinal source and, hence, could not originate from Storm Guchol to propagate to our mooring site because Storm Guchol passed equatorward of the mooring site. However, the passage of tropical storm Haikui over the Philippines around day 310 had such a large extended storm circulation that its northern edges swept over Taiwan. It is hypothesized that the large extended circulation can explain the observation of modest $T^f$ at our mooring site around day 310 (Fig. 4b; second largest $T^f$ in the record).

The above considerations suggest a dominance of planetary vorticity over local mesoscale eddy vorticity. The latter may play some role and are calculated to contribute $(0.03\text{-}0.1)f$ in the upper 1000 m in the area (Morozov and Velarde, 2008). However, all periods of enhanced $T^f$ in the present observational record can be explained from planetary vorticity effects that vary with latitude. It is noted that the latitudinal extent O(1000 km) of large atmospheric disturbances cause a 5% variation in local planetary vorticity.

While long (transient) ocean surface waves may be a first indicator ahead of a cyclone, the fast response in deep-sea- $T^f$ prior to current variations bears similarity to observations from shallow seas where a two-layer stratification exists (Millot and Crépon, 1981; Maas and van Haren, 1987). Although the precise mechanism for energy transfer between barotropic and baroclinic near-inertial IGW is still not known and requires future modelling efforts, our observations suggest a local generation of the latter, possibly after reflection at the underwater topography as has been suggested at the coastal boundary for shallow seas (Millot and Crepon, 1981). The result are IGW with unusually large (>200 m vertical) isothermal excursions at near-inertial frequencies.

The >200 m large vertical near-inertial motions may explain the observed large-scale convective turbulence character, and which dominate over shear-induced motions for brief periods of time and differently observed at OBS sites 500 to 1000 m away. Such turbulent convection is expected to reach the seafloor and affect sediment resuspension locally for several hours. The turbulence duration may be concluded from the OBS-C of which values are enhanced for several hours during the largest convection around day 258. OBS-C waveforms may contain both translational and rotational ground motions (Yang et al., 2018). These observations cannot be attributed to seismic earthquake activity or infra-gravity ocean surface wave activity because here we studied much longer period signals, and also because energy propagated with much slower apparent speeds $O(0.1)$ m s$^{-1}$. Although further studies are required, we propose internal-wave induced turbulence as an additional process for cyclones to generate ground motions on the seafloor.

Our study shows that vigorous turbulence above deep sloping topography nearly doubled when inertial waves broke. It is suggested that quantification of inertial, besides tidal, turbulence processes may be implemented in the parametrization of models on sediment resuspension and benthic life above such topography. This may help improving our understanding of deep-ocean transport processes.

**Acknowledgments**

We thank captains and crew members of the R/V Ocean Researcher 1 (aka OR1) and Ocean Researcher 3 (aka OR3). We thank M. Laan for his ever-lasting thermistor efforts. We benefitted from discussions with Dr. R. Fongngern on suspension sediments during the cruise. We thank Dr. B-Y. Kuo, Dr. Ching-Ren Li, and the OBS team at IES for providing the OBS instruments. We thank the Institute of Oceanography, National Taiwan University for providing multibeam data. The TerraMODIS photo of Typhoon Talim is from NASA. Data associated with the present paper are available from https://doi.pangaea.de/10.1594/PANGAEA.897657.

**Appendix A**

Internal wave motions can have a wide variety of different amplitudes, frequencies and intensities, which obscure the study of particular processes. Sometimes, motions at particular frequencies dominate, which facilitates process studies. In the example given in Fig. 7 inertial motions with vertical isopycnals prevail, while in Fig. 8 semidiurnal tidal motions dominate with stronger stratification. This was made clear in the magnifications of Fig. 9.

An example of a mix of large energy-containing low-frequency internal wave motions is given in Fig. A1, and which occurred around the time of passage of Storm Guchol equatorward of Green Island. During this period of time inertial motions are weak, but the semidiurnal tidal motions are superposed on diurnal tidal motions, while the stratification is relatively horizontal (Fig. A1c, compare with Fig. 7c where it is more vertically oriented). While the temperature variations seem more chaotic than in Fig. 8, the motions do resemble that semidiurnal-dominated image much more than the inertial-dominated period in Fig. 7. The exception are the larger mean turbulence values ($[<\varepsilon>] = 0.8\pm0.4\times10^{-7}$ m$^2$ s$^{-3}$, $[<K_z>] = 4\pm2\times10^{-2}$ m$^2$ s$^{-1}$, while $<[N]> = 6\pm1\times10^{-4}$ s$^{-1}$) which approach, but do not reach, those during the inertial period of Fig. 7. The largest overturning in Fig. A1 occurs in two intense episodes one day apart.

**References**

Briscoe, M.G., 1975. Preliminary results from the trimoored internal wave experiment (IWEX). Journal of Geophysical Research 80, 3872-3884.

Chang, S.W., 1985. Deep ocean response to hurricanes revealed by an ocean model with a free surface. Journal of Physical Oceanography 15, 1847-1858.

Chi, W.-C., Chen, W.-J., Kuo, B.-Y., Dolenc, D., 2010a. Seismic monitoring of western Pacific typhoons. Marine Geophysical Research 31, 239-251.

Chi, W.-C. et al., 2010b. Seismological report on the 2006 Typhoon Shanshan that lits up seismic stations along its way. Seismological Research Letters 81, 592-596, doi: 10.1785/gssrl.81.4.592.

Cimatoribus, A.A., van Haren, H., Gostiaux, L., 2014. Comparison of Ellison and Thorpe scales from Eulerian ocean temperature observations. Journal of Geophysical Research 119, 7047-7065, doi:10.1002/2014JC010132.

Cox, C., Deaton, T., Webb, S., 1984. A deep-sea differential pressure gauge. Journal of Atmospheric and Oceanic Technology 1, 237-246.

Cuypers, Y., Le Vaillant, X., Bouruet-Aubertot, P., Vialard, J., McPhaden, M., 2013. Tropical storm-induced near-inertial internal waves during the Cirene experiment: Energy fluxes and impact on vertical mixing. Journal of Geophysical Research 118, 358-380.

D'Asaro, E.A., Lien, R.-C., 2000. The wave-turbulence transition for stratified flows. Journal of Physical Oceanography 30, 1669-1678.

D'Asaro, E.A. et al., 2014. Impact of typhoons on the ocean in the Pacific. Bulletin of the American Meteorological Society 95, 1405-1418.

Davy, C., Barruol, G., Fontaine, F.R., Sigloch, K., Stutzmann, E., 2014. Tracking major storms from microseismic and hydroacoustic observations on the seafloor. Geophysical Research Letters 41, 8825-8831, doi:10.1002/2014GL062319.

Dillon, T.M., 1982. Vertical overturns: a comparison of Thorpe and Ozmidov length scales. Journal of Geophysical Research 87, 9601-9613.
21

**Figure 1**. (Background photo) Position and extent of Cat. 4 (Saffir-Simpson scale) tropical cyclone Typhoon Talim on 14 September 2017, 02:15 UTC which is yearday 256.0938. Northern Taiwan is obscured by Typhoon Talim's SW-side clouds (TerraMODIS image, NASA). The pink square indicates the observational area, which is magnified in b and c. (a) Best-guessed track from http://www.metoc.navy.mil/jtwc/jtwc.html portal, see also http://agora.ex.nii.ac.jp/digital-typhoon/summary/wnp/s/201718.html.en. blue, yellow, red and purple filled circles indicate tropical depression, tropical storm, tropical cyclone and extratropical cyclone, respectively. Yearday 256.0938 position in circle. The pink square refers to b and c. (b) Map around Green Island of 100x100 m smoothed Multibeam bathymetry with solid contours indicating 0, 3000, 3100 and 3200 m. The rectangle indicates the area of Fig. 2 with mooring locations. (c) Same as b., but for inferred 2D seafloor slope. The colored range indicates internal tide slopes varying with stratification; white indicates the range of near-inertial slopes.

**Figure 2**. Mooring site detail. Seismic recorders OBS-A,B,C are indicated by solid purple dots, temperature T-sensor mooring by the red dot. Note the different color scale of the detailed bathymetry compared to Fig. 1b. Topography contours are drawn every 50 m of elevation.

**Figure 3**. Conservative Temperature-density anomaly relationship, from the lower 500 m of CTD-data obtained within 1 km from the mooring site.

**Figure 4**. Overview records from OBS, current meter CM and T-sensors. (a) Time series of entire 10 months record of cross-slope current component observed at 208 m above the seafloor. Positive values are off-slope. In blue original data, in red near-inertial band-pass filtered record using a double elliptic filter with cut-off frequencies at 0.85f and 1.25f, f the local inertial frequency (superscript f). The grey shading indicates the period of d.-f. (b) As a., but for CM-along slope current component, positive is poleward. (c) As a., but for CM-

T. (d) September period showing $T^f$ from c. together with unfiltered upper (z = -2937 m, green) and lower (z = -3137 m, black) T-sensor data. Note the different vertical scale compared to c. The coloured shading denotes the time of passage of Storm Guchol in the Strait of Luzon (blue), and the intensity of Typhoon Talim following the coding in Fig. 1a. The dark-red is the period when Typhoon Talim reached Cat. 4, which coincided with the period of nearest approach (eye still 600 km from the mooring site) and after which it veered to the northeast. (e) As d. but for normalized low-frequency OBS-p waveform for stations A, B, C in light-blue, blue, red, respectively. (f) As d. but for logarithm of 200-m vertically averaged turbulence estimates. In black turbulent diffusivity (scale to left), in purple turbulence dissipation rate (scale to right).

**Figure 5**. Calculations of directionality of propagation of low-frequency internal waves. (a) Hodograph of 4 days of band-pass filtered near-inertial horizontal current ellipses from CM-data at z = -2936 m, between 11 and 15 September during the approach of Typhoon Talim. (b) As a., but for 15-19 September. (c) Back-azimuth of summed relative power in gridded bins (colours) from frequency-wavenumber array analysis of low-pass filtered OBS-Z arrayed data for the period between 11 and 19 September. Circle grids indicate the slowness in s km$^{-1}$, so that a value of 2000 implies a propagation velocity of 0.5 m s$^{-1}$.

**Figure 6**. Spectra of kinetic energy and temperature variance. (a) Nearly raw (3 degrees of freedom) CM-kinetic energy and CM-T-spectra from z = -2936 m for the 10-day period starting at day 244 (2 September). In black kinetic energy, in blue T-variance. $M_2$ denotes the semidiurnal lunar tidal frequency, N the large 100-m-scale buoyancy frequency. (b) As a., but for 10-day period starting at day 254 (12 September).

**Figure 7**. Three-day detail of large inertial motion period in mid-September 2017. (a) Time series of horizontal component amplitudes OBS-C from moorings B (positive absolute;

blue), C (negative absolute; red). (b) Time-depth series of Conservative (~potential) Temperature from T-sensors. The purple bar indicates the mean local buoyancy period, the white bar indicates the local inertial period, and the black bar indicates the semidiurnal (thick) and diurnal (thick and thin) periods. The sea-floor is at the horizontal axis. (c) Time-depth series of logarithm of local buoyancy frequency from reordered profiles of b. (d) Time series of logarithm of 200-m vertically averaged turbulence dissipation rates.

**Figure 8**. As Fig. 7 with identical scales, but for a semidiurnal tidal dominated period a week before.

**Figure 9**. Six-hour magnifications, with black contours are drawn every 0.02°C, of: (a) Fig. 7b. (b) Fig. 8b. Note the different colour scales, the twice larger buoyancy period (purple bar) in a. compared to that of b. and the different but large >200-m overturning.

**Figure 10**. Five months (yeardays 160 to 320) averaged spectra. Data from all independent temperature sensors are sub-sampled to once per 10 s for computational reasons. (a) Heavily smoothed (1000 dof, degrees of freedom) mean temperature variance spectrum. $M_2$ denotes the semidiurnal tidal frequency, f the local inertial frequency and N the large-100-m-scale mean buoyancy frequency with $N_{min}$ its two-hour averaged minimum value. The spectra are compared with slopes of turbulence inertial subrange power law $\sigma^{-5/3}$ and canonical internal wave slope of $\sigma^{-2}$. (b) Smoothed (100 dof) coherence spectra from all possible pairs of T-sensors at different vertical separations. $N_{max}$ denotes the small(2-m)-scale maximum buoyancy frequency.

**Figure A1**. As Fig. 7 with identical scales, but for a semidiurnal and diurnal tidal dominated period around the time of Storm Guchol passing equatorward of the mooring.

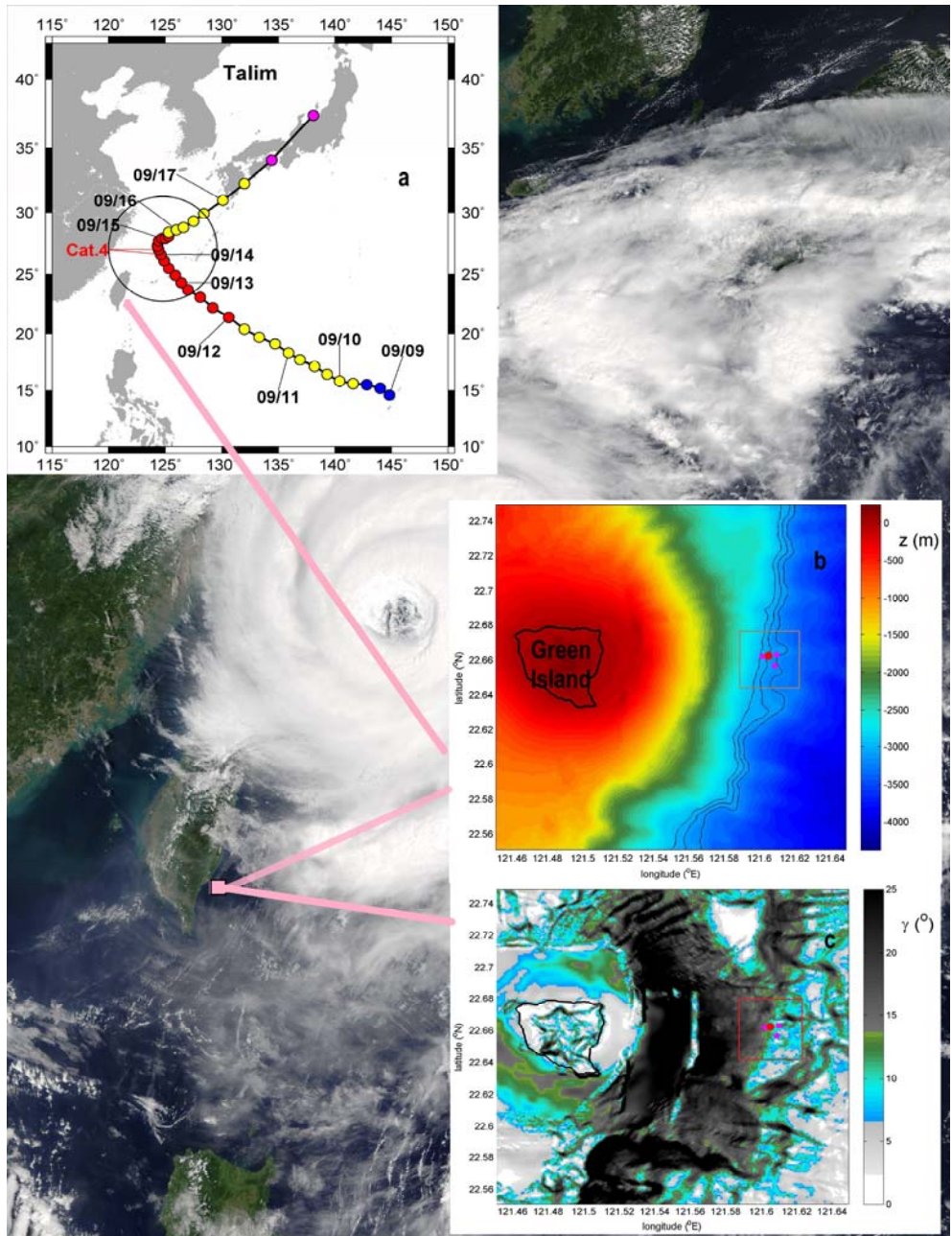

**Figure 1**. (Background photo) Position and extent of Cat. 4 (Saffir-Simpson scale) tropical cyclone Typhoon Talim on 14 September 2017, 02:15 UTC which is yearday 256.0938. Northern Taiwan is obscured by Typhoon Talim's SW-side clouds (TerraMODIS image, NASA). The pink square indicates the observational area, which is magnified in b and c. (a) Best-guessed track from http://www.metoc.navy.mil/jtwc/jtwc.html portal, see also http://agora.ex.nii.ac.jp/digital-typhoon/summary/wnp/s/201718.html.en. blue, yellow, red and purple filled circles indicate tropical depression, tropical storm, tropical cyclone and extratropical cyclone, respectively. Yearday 256.0938 position in circle. The pink square refers to b and c. (b) Map around Green Island of 100x100 m smoothed Multibeam bathymetry with solid contours indicating 0, 3000, 3100 and 3200 m. The rectangle indicates the area of Fig. 2 with mooring locations. (c) Same as b., but for inferred 2D seafloor slope. The colored range indicates internal tide slopes varying with stratification; white indicates the range of near-inertial slopes.

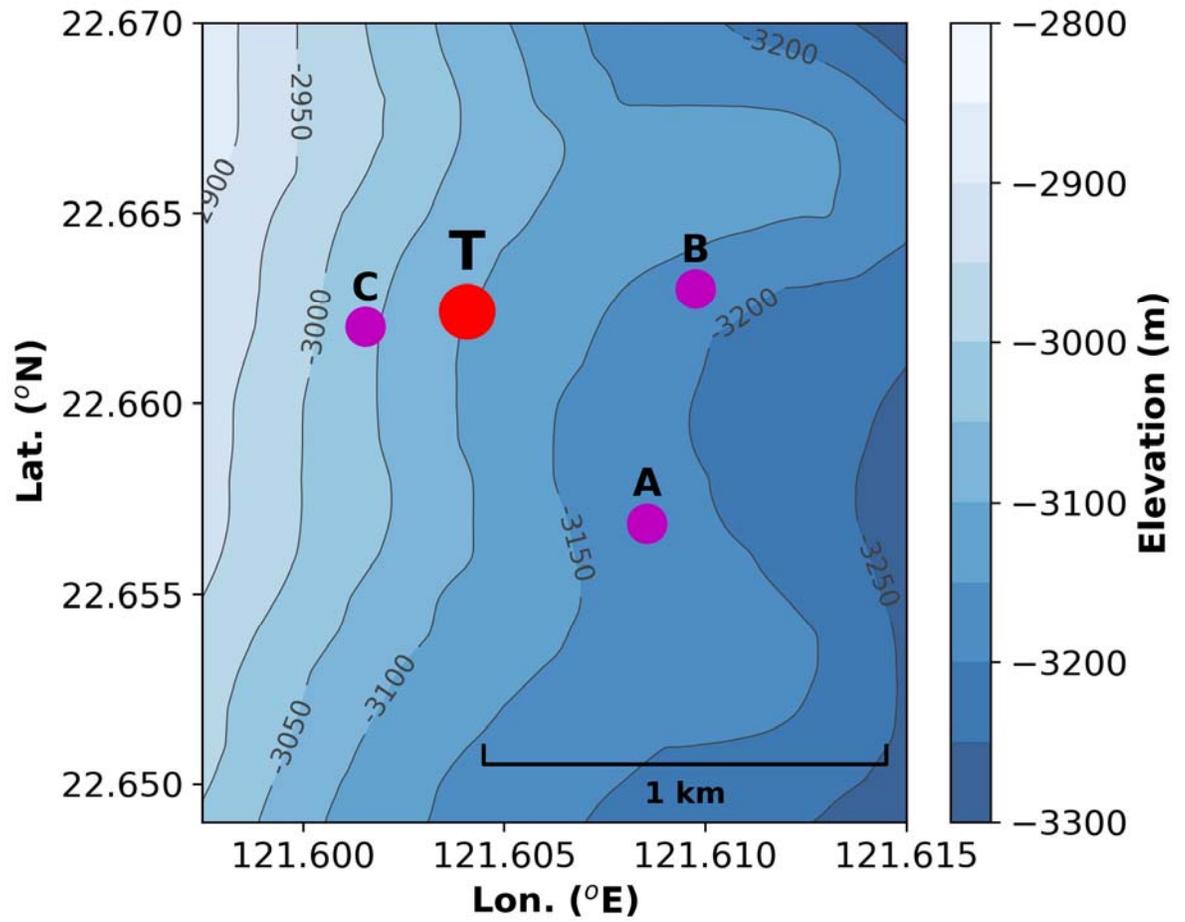

**Figure 2**. Mooring site detail. Seismic recorders OBS-A,B,C are indicated by solid purple dots, temperature T-sensor mooring by the red dot. Note the different color scale of the detailed bathymetry compared to Fig. 1b. Topography contours are drawn every 50 m of elevation.

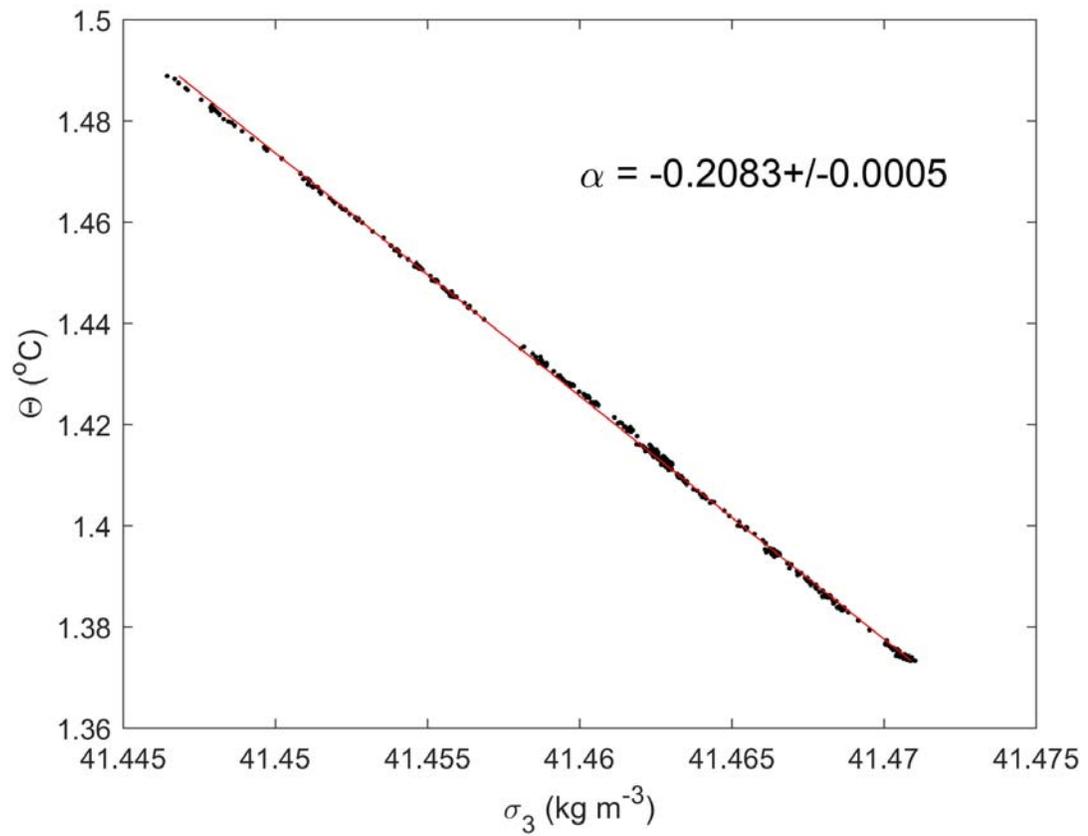

**Figure 3**. Conservative Temperature-density anomaly relationship, from the lower 500 m of CTD-data obtained within 1 km from the mooring site.

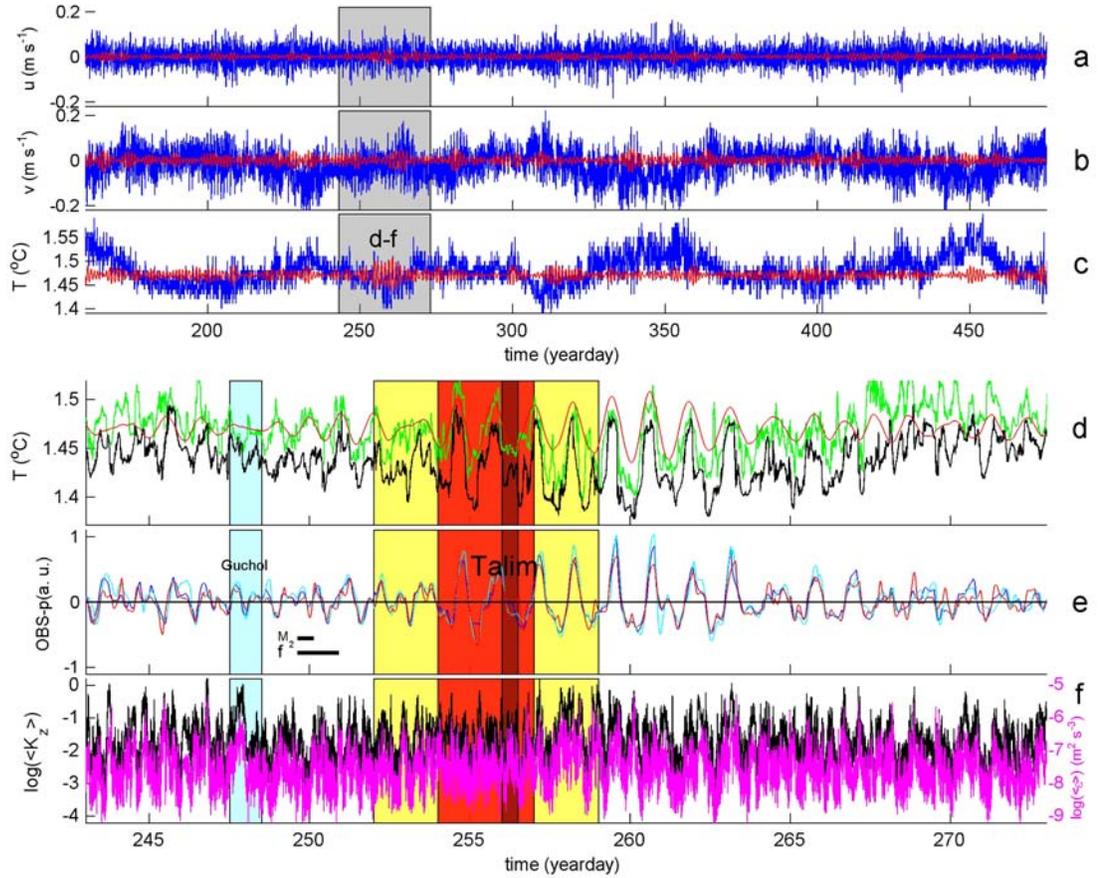

**Figure 4**. Overview records from OBS, current meter CM and T-sensors. (a) Time series of entire 10 months record of cross-slope current component observed at 208 m above the seafloor. Positive values are off-slope. In blue original data, in red near-inertial band-pass filtered record using a double elliptic filter with cut-off frequencies at 0.85f and 1.25f, f the local inertial frequency (superscript f). The grey shading indicates the period of d.-f. (b) As a., but for CM-along slope current component, positive is poleward. (c) As a., but for CM-T. (d) September period showing $T^f$ from c. together with unfiltered upper ($z = -2937$ m, green) and lower ($z = -3137$ m, black) T-sensor data. Note the different vertical scale compared to c. The coloured shading denotes the time of passage of Storm Guchol in the Strait of Luzon (blue), and the intensity of Typhoon Talim following the coding in Fig. 1a. The dark-red is the period when Typhoon Talim reached Cat. 4, which coincided with the period of nearest approach (eye still 600 km from the mooring site) and after which it veered to the northeast. (e) As d. but for normalized low-frequency OBS-p waveform for stations A, B, C in light-blue, blue, red, respectively. (f) As d. but for logarithm of 200-m vertically averaged turbulence estimates. In black turbulent diffusivity (scale to left), in purple turbulence dissipation rate (scale to right).

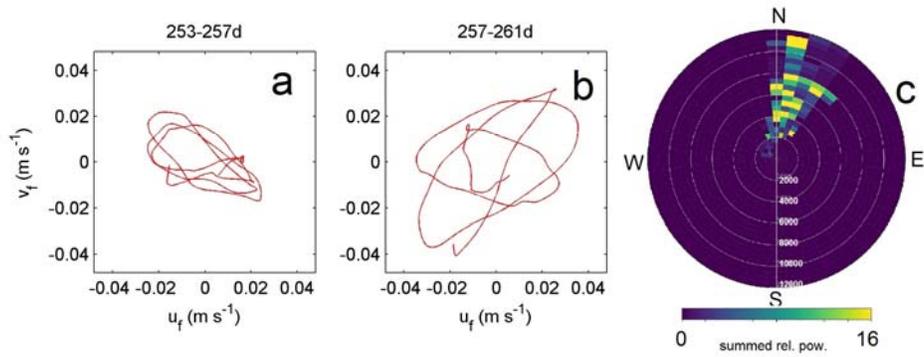

**Figure 5**. Calculations of directionality of propagation of low-frequency internal waves. (a) Hodograph of 4 days of band-pass filtered near-inertial horizontal current ellipses from CM-data at $z = -2936$ m, between 11 and 15 September during the approach of Typhoon Talim. (b) As a., but for 15-19 September. (c) Back-azimuth of summed relative power in gridded bins (colours) from frequency-wavenumber array analysis of low-pass filtered OBS-Z arrayed data for the period between 11 and 19 September. Circle grids indicate the slowness in s km$^{-1}$, so that a value of 2000 implies a propagation velocity of 0.5 m s$^{-1}$.

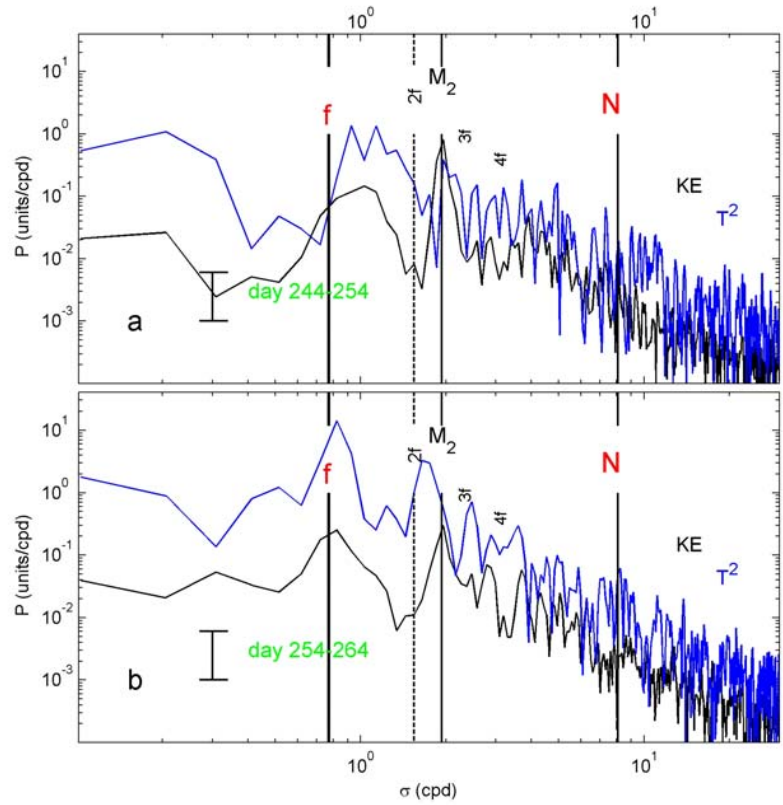

**Figure 6**. Spectra of kinetic energy and temperature variance. (a) Nearly raw (3 degrees of freedom) CM-kinetic energy and CM-T-spectra from z = -2936 m for the 10-day period starting at day 244 (2 September). In black kinetic energy, in blue T-variance. $M_2$ denotes the semidiurnal lunar tidal frequency, N the large 100-m-scale buoyancy frequency. (b) As a., but for 10-day period starting at day 254 (12 September).

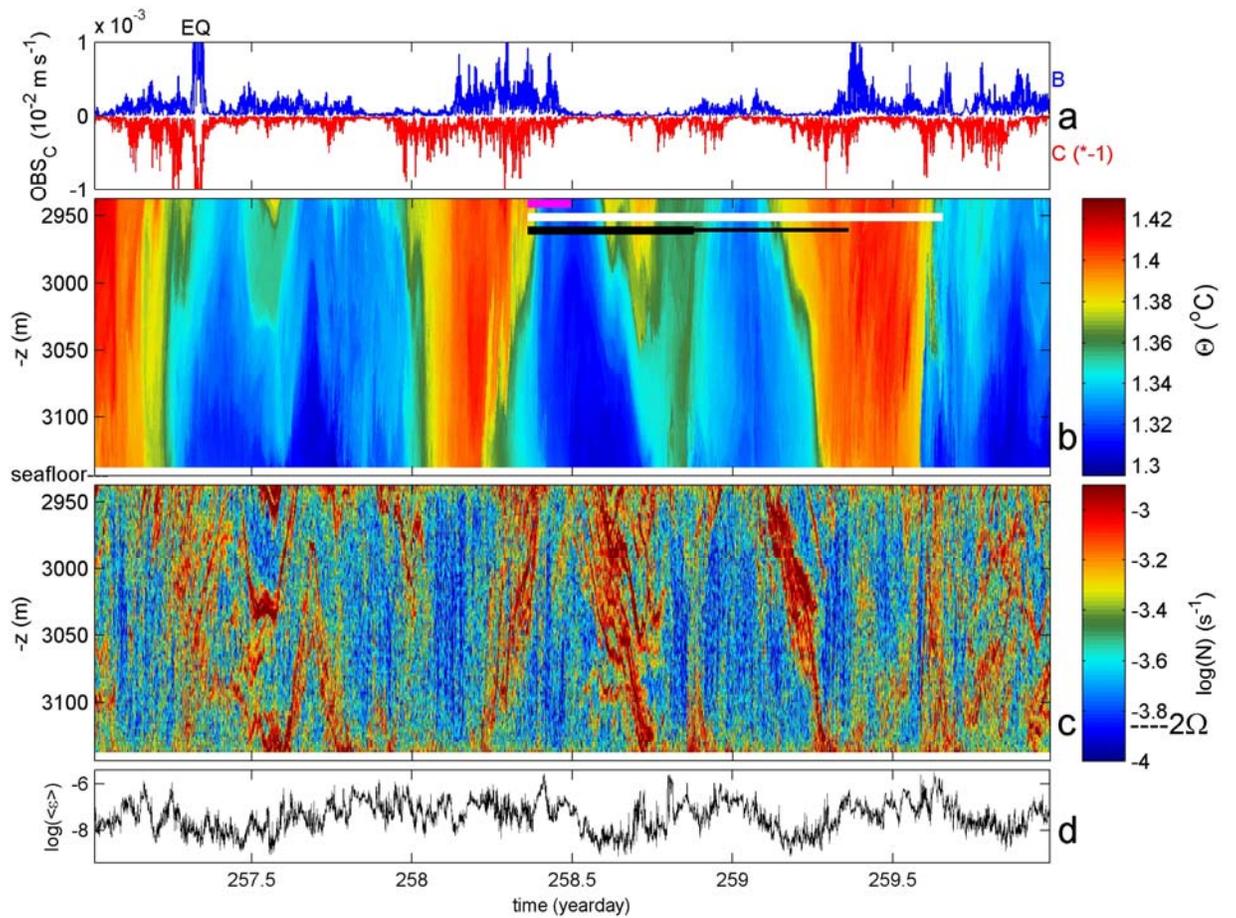

**Figure 7**. Three-day detail of large inertial motion period in mid-September 2017. (a) Time series of horizontal component amplitudes OBS-C from moorings B (positive absolute; blue), C (negative absolute; red). (b) Time-depth series of Conservative (~potential) Temperature from T-sensors. The purple bar indicates the mean local buoyancy period, the white bar indicates the local inertial period, and the black bar indicates the semidiurnal (thick) and diurnal (thick and thin) periods. The sea-floor is at the horizontal axis. (c) Time-depth series of logarithm of local buoyancy frequency from reordered profiles of b. (d) Time series of logarithm of 200-m vertically averaged turbulence dissipation rates.

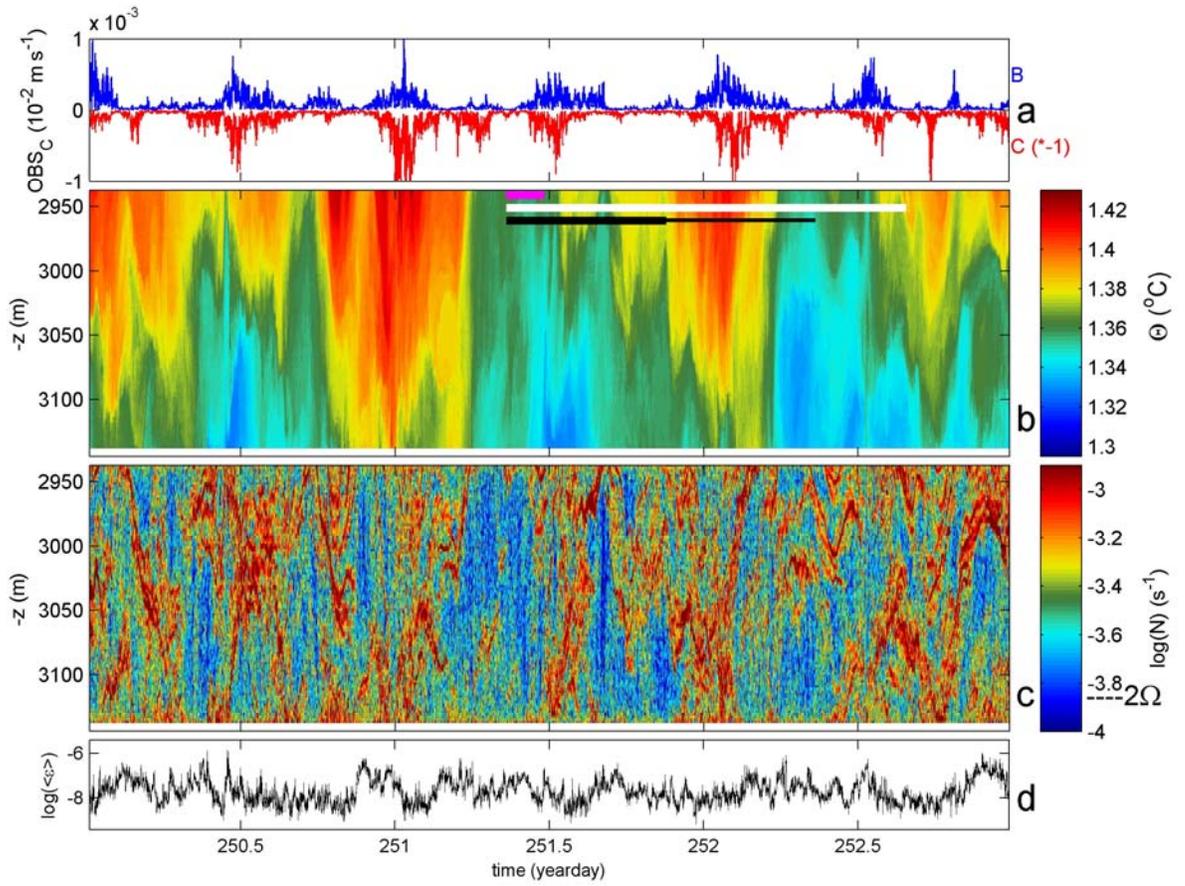

**Figure 8**. As Fig. 7 with identical scales, but for a semidiurnal tidal dominated period a week before.

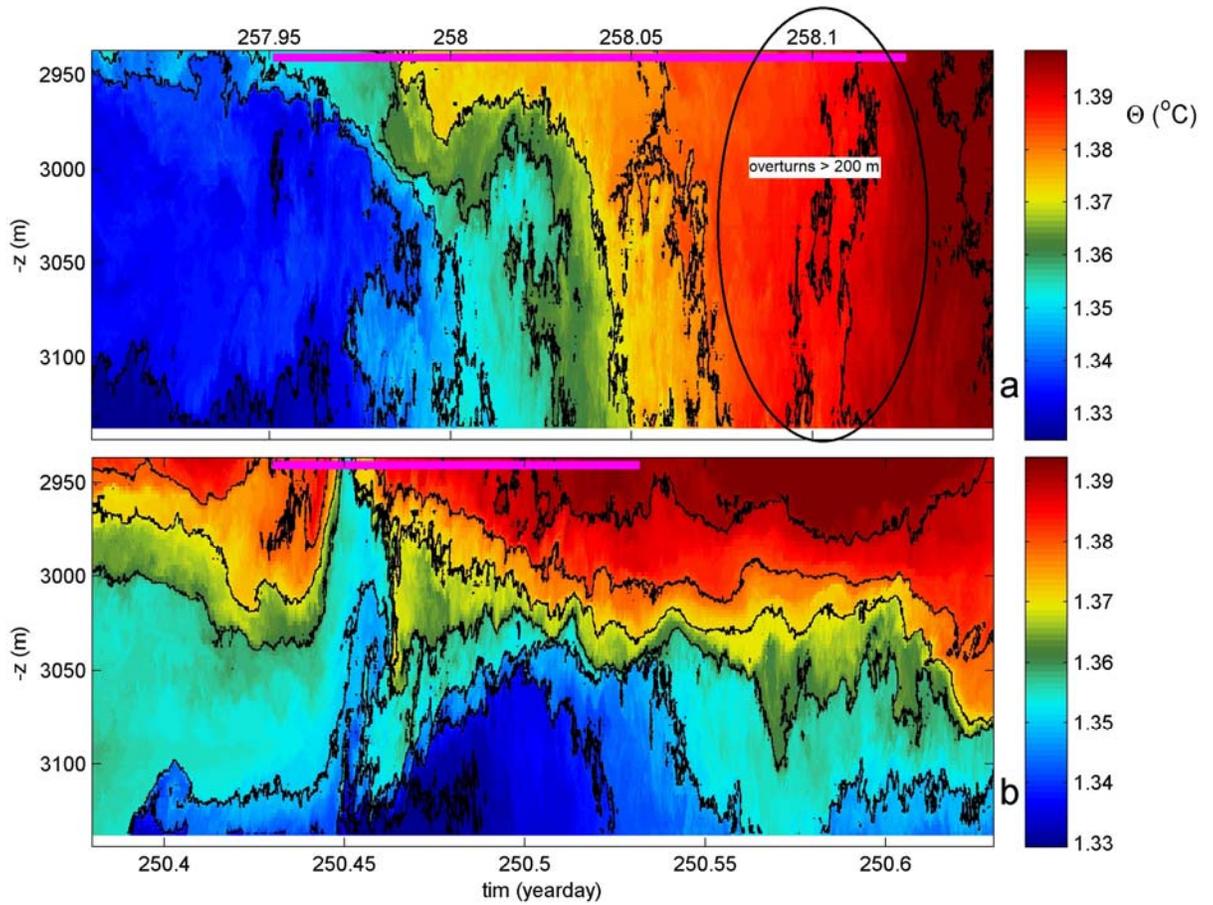

**Figure 9**. Six-hour magnifications, with black contours are drawn every 0.02°C, of: (a) Fig. 7b. (b) Fig. 8b. Note the different colour scales, the twice larger buoyancy period (purple bar) in a. compared to that of b. and the different but large >200-m overturning.

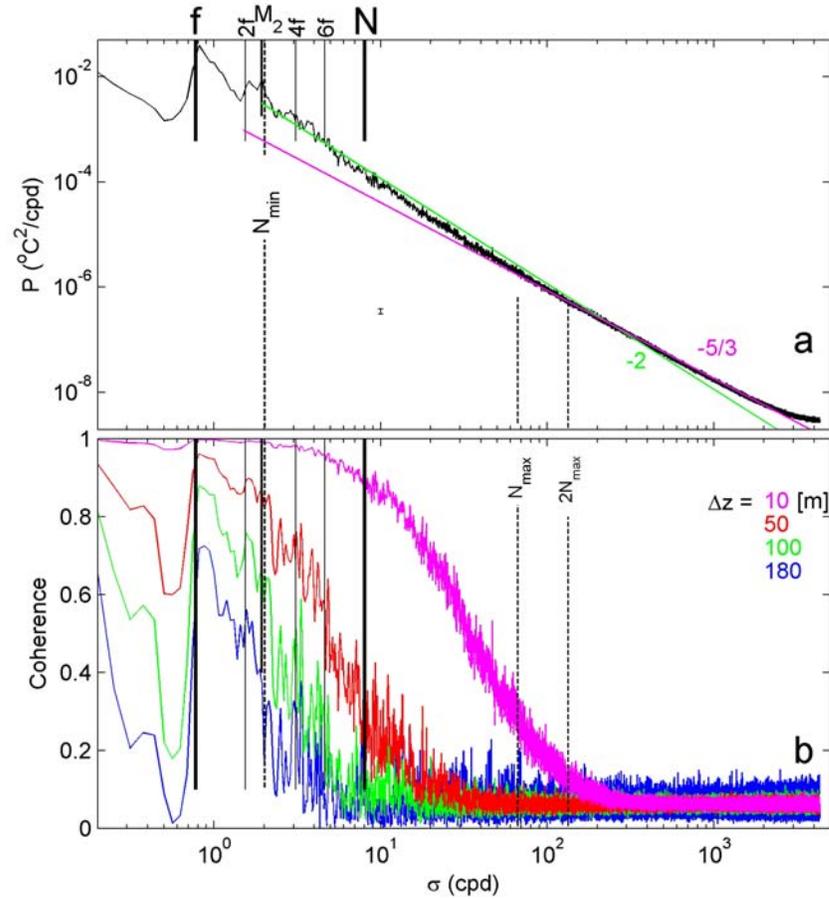

**Figure 10**. Five months (yeardays 160 to 320) averaged spectra. Data from all independent temperature sensors are sub-sampled to once per 10 s for computational reasons. (a) Heavily smoothed (1000 dof, degrees of freedom) mean temperature variance spectrum. $M_2$ denotes the semidiurnal tidal frequency, f the local inertial frequency and N the large-100-m-scale mean buoyancy frequency with $N_{min}$ its two-hour averaged minimum value. The spectra are compared with slopes of turbulence inertial subrange power law $\sigma^{-5/3}$ and canonical internal wave slope of $\sigma^{-2}$. (b) Smoothed (100 dof) coherence spectra from all possible pairs of T-sensors at different vertical separations. $N_{max}$ denotes the small(2-m)-scale maximum buoyancy frequency.

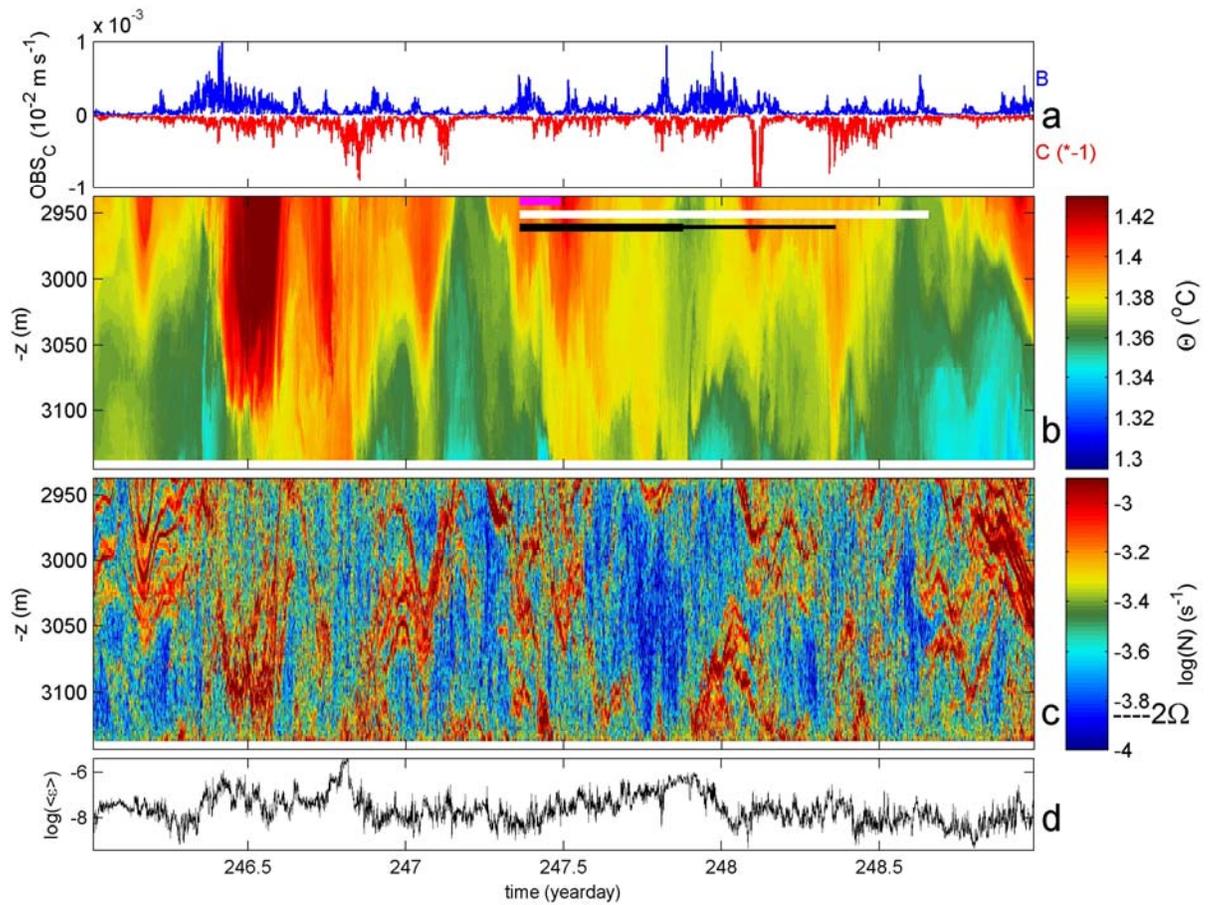

**Figure A1.** As Fig. 7 with identical scales, but for a semidiurnal and diurnal tidal dominated period around the time of Storm Guchol passing equatorward of the mooring.